\documentclass[sn-mathphys-num]{sn-jnl}

\usepackage{graphicx}%
\usepackage{multirow}%
\usepackage{amsmath,amssymb,amsfonts}%
\usepackage{amsthm}%

\usepackage{mathrsfs}%
\usepackage[title]{appendix}%
\usepackage{xcolor}%
\usepackage{textcomp}%
\usepackage{manyfoot}%
\usepackage{booktabs}%
\usepackage{algorithm}%
\usepackage{algorithmicx}%
\usepackage{algpseudocode}%
\usepackage{listings}%
\usepackage{colortbl}

\theoremstyle{thmstyleone}%
\theoremstyle{thmstyletwo}%

\theoremstyle{thmstylethree}%
\begin{document}

\title[Article Title]{Enhancing nuclear cross-section predictions with deep learning: the DINo algorithm}

\author*[1]{\fnm{Lévana} \sur{Gesson}}\email{levana.gesson@iphc.cnrs.fr}
\author[1]{\fnm{Greg} \sur{Henning}}\email{greg.henning@iphc.cnrs.fr}
\author[1]{\fnm{Jonathan} \sur{Collin}}\email{jonathan.collin@iphc.cnrs.fr}
\author[1]{\fnm{Marie} \sur{Vanstalle}}\email{marie.vanstalle@iphc.cnrs.fr}

\affil[1]{\orgname{Universite de Strasbourg, CNRS, IPHC UMR 7178}, \orgaddress{\city{Strasbourg}, \postcode{67000}, \country{France}}}


\abstract{Accurate modeling of nuclear reaction cross-sections is crucial for applications such as hadron therapy, radiation protection, and nuclear reactor design. Despite continuous advancements in nuclear physics, significant discrepancies persist between experimental data and theoretical models such as TENDL (TALYS-based Evaluated Nuclear Data Library), and ENDF/B (Evaluated Nuclear Data File). These deviations introduce uncertainties in Monte Carlo simulations widely used in nuclear physics and medical applications. In this work, DINo (Deep learning Intelligence for Nuclear reactiOns) is introduced as a deep learning-based algorithm designed to improve cross-section predictions by learning correlations between charge-changing and total cross-sections.\\

Trained on the TENDL-2021 dataset and validated against experimental data from the EXFOR database, DINo demonstrates a significant improvement in predictive accuracy over conventional nuclear models. The results show that DINo systematically achieves lower $\chi^2$ values compared to TENDL-2021 across multiple isotopes, particularly for proton-induced reactions on a $^{12}$C target, a key material in hadron therapy. Specifically, for $^{11}$C production, DINo reduces the discrepancy with experimental data by $\sim 28\%$ compared to TENDL-2021. Additionally, DINo provides improved predictions for other relevant isotopes produced, such as $^4$He, $^6$Li, $^9$Be, and $^{10}$B, which play a crucial role in modeling nuclear fragmentation processes.\\

By leveraging neural networks, DINo offers fast cross-section predictions, making it a promising complementary tool for nuclear reaction modeling. However, the algorithm's performance evaluation is sensitive to the availability of experimental data, with increased uncertainty in sparsely measured energy ranges. Future work will focus on refining the model through data augmentation, expanding its applicability to other reaction channels, and integrating it into Monte Carlo transport codes for real-time nuclear data processing. These advances could significantly enhance predictive capabilities in nuclear physics, and medical applications.}

\keywords{Nuclear cross-sections, Deep learning, Hadron therapy, Monte Carlo simulations, Proton-induced reactions}



\maketitle

\section{Introduction}\label{Introduction}

Accurate modeling of nuclear reactions is essential for numerous applications, including hadron therapy, nuclear reactor design, and radiation protection. However, despite continuous advancements in nuclear physics, persistent discrepancies between experimental data and theoretical nuclear models remain a significant challenge, particularly in particle therapy, where precise cross-section predictions directly impact treatment planning and dosimetry \cite{Battistoni2016}.

Traditionally, nuclear models such as TENDL (TALYS-based Evaluated Nuclear Data Library) \cite{TENDL} and ENDF/B (Evaluated Nuclear Data File) \cite{ENDF} rely on a combination of theoretical calculations and experimental benchmark. However, these models often exhibit systematic deviations from measured cross-sections. Consequently, Monte Carlo (MC) simulations—widely used for predicting nuclear interactions in medical and nuclear physics applications—inherit of this biaised data.

A conventional approach to mitigating these inaccuracies involves integrating experimental cross-section measurements to refine Monte Carlo simulations \cite{FOOT}. While this method enhances predictive reliability, it remains dependent on the availability of high-quality experimental data, which is often limited for certain isotopes and energy ranges. An alternative strategy is to directly enhance the predictive capabilities of nuclear models by incorporating machine learning techniques capable of capturing complex patterns within nuclear data. Such attempts have been investigate in a similar approach for inelastic neutron scattering cross-sections \cite{Henning}.\\

In this work, initially developed in the PhD thesis \cite{Gesson}, DINo (Deep learning Intelligence for Nuclear reactiOns) is introduced as a deep learning-based algorithm designed to address these challenges. DINo leverages neural networks to learn the underlying correlations between partial nuclear reaction cross-sections and total cross-sections, thereby improving the accuracy of nuclear reaction predictions beyond what is achievable with conventional nuclear models. Specifically, DINo is trained using TENDL-2021 nuclear data, to predict as a starting point of the total cross-sections for proton-induced nuclear reactions on a $^{12}$C target. The choice of protons is motivated by their prominent role in hadron therapy, where accurate cross-section estimations are crucial for dose calculations and treatment optimization.

One of DINo’s key advantages is its ability to generalize beyond the training dataset, allowing for accurate predictions even in energy regions lacking direct experimental measurements. By systematically comparing DINo’s predictions against both experimental data from EXFOR and TENDL-2021 model outputs, the potential of deep learning is evaluated to reduce systematic uncertainties in nuclear reaction modeling.

\section{Material and methods}\label{MatMet}

\subsection{TENDL-2021 nuclear model and EXFOR experimental data}
TENDL (TALYS-based Evaluated Nuclear Data Library) \cite{TENDL} is a nuclear data library produced at NRG Petten. Since 2015, TENDL is mainly developed at PSI and the IAEA (Nuclear Data Section). While originally based on the TALYS theoretical nuclear reaction code, the main version distributed incorporates data from the other evaluator ENDF/B (Evaluated Nuclear Data File) library \cite{ENDF}, first distributed with ENDF/B-VII.0 and still available in ENDF/B-VIII.0. This remains the case for the reaction of interest.

Nuclear data libraries like ENDF/B and TENDL provide essential information on nuclear reactions, including cross-sections, decay data, and other fundamental nuclear properties. Each new release of TENDL integrates refinements in theoretical modeling, updated experimental data, and corrections based on benchmarking against experimental results and user feedback.

In this study, protons were selected due to their extensive coverage in nuclear data libraries. To ensure a uniform energy step distribution and standardize the ENDF6 \cite{ENDF6_manual}, an algorithm developed by J. Collin (private communication) was applied. \\

EXFOR (Experimental Nuclear Reaction Data) \cite{EXFOR} is an international database of experimental nuclear reaction data, maintained by the International Network of Nuclear Reaction Data Centers (NRDC) with support from the International Atomic Energy Agency (IAEA).

The database compiles measured nuclear cross-sections, energy spectra, and angular distributions obtained from worldwide experiments. It standardizes data formats for easy integration into nuclear models. EXFOR gathers data from a wide range of experimental facilities, including accelerators, reactors, and other nuclear research facilities. Data contributors can include universities, research institutes, and national laboratories. The data are publicly accessible and can be accessed through various channels, including the EXFOR database website, which provides search and retrieval functionalities. Users can search for specific reactions, isotopes, energy ranges, and other parameters to retrieve relevant experimental data.

\begin{figure*}[!htbp]
\begin{center}
\includegraphics[scale=.6]{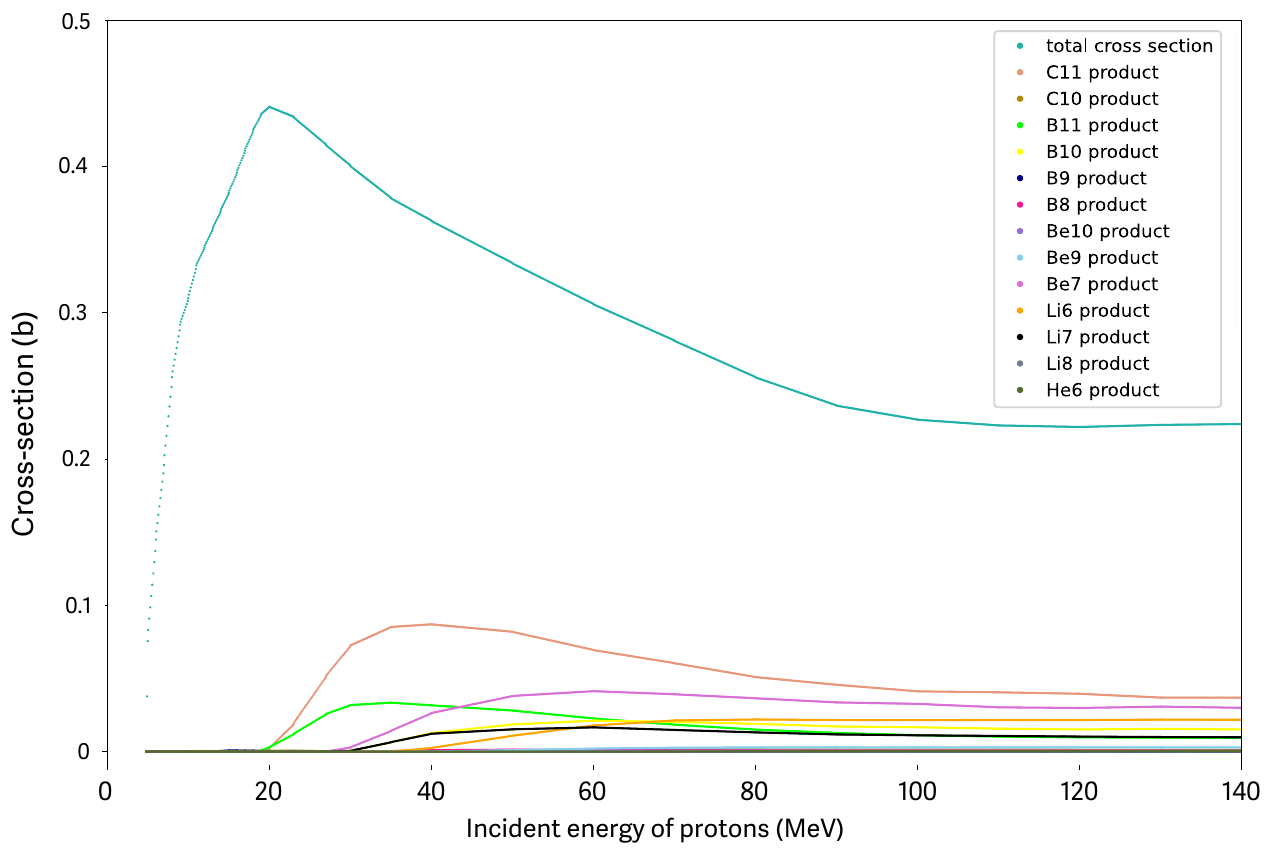}
\caption{\label{fig:TENDL} TENDL-2021 model of total and partial cross-sections of a proton beam on a $^{12}$C target.}
\end{center}
\end{figure*}

Since proton-induced reactions are well-documented in EXFOR, this dataset was chosen to evaluate and validate our deep learning model, and more especially a proton beam on a $^{12}$C target, providing a direct comparison between experimental and theoretical nuclear cross-sections.
Figure \ref{fig:TENDL} shows the total and partial cross-sections for incident protons on a $^{12}$C target from the TENDL-2021 nuclear model. 

Firstly, $^{12}$C is often referred to as main constituent of "tissue-equivalent" material, meaning it has similar nuclear properties to biological tissues. This characteristic makes it an ideal target for simulating interactions that occur in human tissues during therapeutic procedures, such as in hadron therapy.

Secondly, the interaction of a proton beam with a $^{12}$C target can produce secondary particles such as protons, neutrons, and fragments of carbon nuclei coming from the target. These secondary particles are crucial for understanding the dosimetry and biological effects of the radiation delivered to the patient. This choice of using a proton projectile on a $^{12}$C target is justified because, in modern nuclear physics experiments that measure cross-sections, such as the FOOT experiment \cite{FOOT}, direct measurement of carbon ion fragmentation in tissue is complex and challenging. This approach, inverse cinematic of carbon fragmentation, leverages the symmetry of nuclear interactions: the fragments produced when a proton hits a $^{12}$C nucleus are similar to those produced when a $^{12}$C ion hits a hydrogen nucleus (which is a major component of human tissue). This makes proton-induced reactions an indirect yet effective method for studying nuclear fragmentation in medical applications.

\subsection{DINo deep learning algorithm}
Deep learning is a subset of machine learning where multi-layered neural networks learn hierarchical representations from data. Unlike traditional methods that require manual feature extraction, deep learning can automatically learn complex relationships between input and output data.
In this work, TensorFlow \cite{Tensorflow} and Keras \cite{Keras} were used to develop a dense neural network (DNN) model trained on the TENDL-2021 dataset.

\subsubsection*{TensorFlow}
TensorFlow \cite{Tensorflow} is an open-source deep learning framework developed by Google that facilitates building and training neural networks. It provides a flexible platform for numerical computations and supports both CPU and GPU acceleration, making it suitable for large-scale machine learning tasks.

\begin{itemize}

\item \textbf{TensorFlow's core} library provides fundamental building blocks for creating computational graphs, managing tensors (multidimensional arrays), and executing operations efficiently. It includes low-level APIs for tensor manipulation and graph construction.

\item \textbf{TFX extends TensorFlow's} capabilities to support end-to-end machine learning workflows. It includes components for data preprocessing, model training, model evaluation, and model serving. TFX is designed for scalable and production-grade machine learning pipelines.

\item \textbf{TensorFlow Serving} is a system for serving machine learning models in production environments. It provides APIs for serving trained models over RESTful endpoints, enabling real-time predictions for applications.

\end{itemize}

\subsubsection*{Keras}
Keras \cite{Keras}, on the other hand, is a high-level neural networks API that simplifies the process of building and training models. It is integrated into TensorFlow as `tf.keras` and offers a user-friendly interface for defining neural network architectures, specifying training parameters, and executing training and evaluation steps.

\begin{itemize}

\item \textbf{Modularity} - Keras allows developers to build neural networks by stacking modular layers. Layers can be easily added, removed, or modified, making it flexible for experimenting with different architectures.

\item Keras provides a wide range of \textbf{pre-built layers} such as dense (fully connected), convolutional, recurrent, dropout, and normalization layers. These layers can be combined to create complex neural network architectures.

\item Keras includes various \textbf{optimizers} (e.g., SGD, Adam, RMSprop) and \textbf{loss functions} (e.g., MSE, categorical cross-entropy) that can be specified during model compilation. This flexibility allows developers to customize training parameters based on the task and dataset. 

\item \textbf{Model Training and Evaluation} - With Keras, training a model involves specifying training data, validation data, batch size, number of epochs, and other parameters. After training, models can be evaluated on test data to assess their performance metrics such as accuracy, precision, and recall.

\end{itemize}

By leveraging the combined power of TensorFlow for efficient computations and Keras for streamlined model building, deep learning models can be created and deployed for a wide range of applications.\\

A dense neural network (DNN), also known as a fully connected neural network, is a fundamental type of artificial neural network (ANN) where each neuron in one layer is connected to every neuron in the adjacent layer. Introduced as a DNN, the DINo algorithm is designed
to predict total cross-sections from charge-changing cross-sections for proton interactions on a $^{12}$C target.
 
\begin{figure*}[!htbp]
\begin{center}
\includegraphics[scale=0.44]{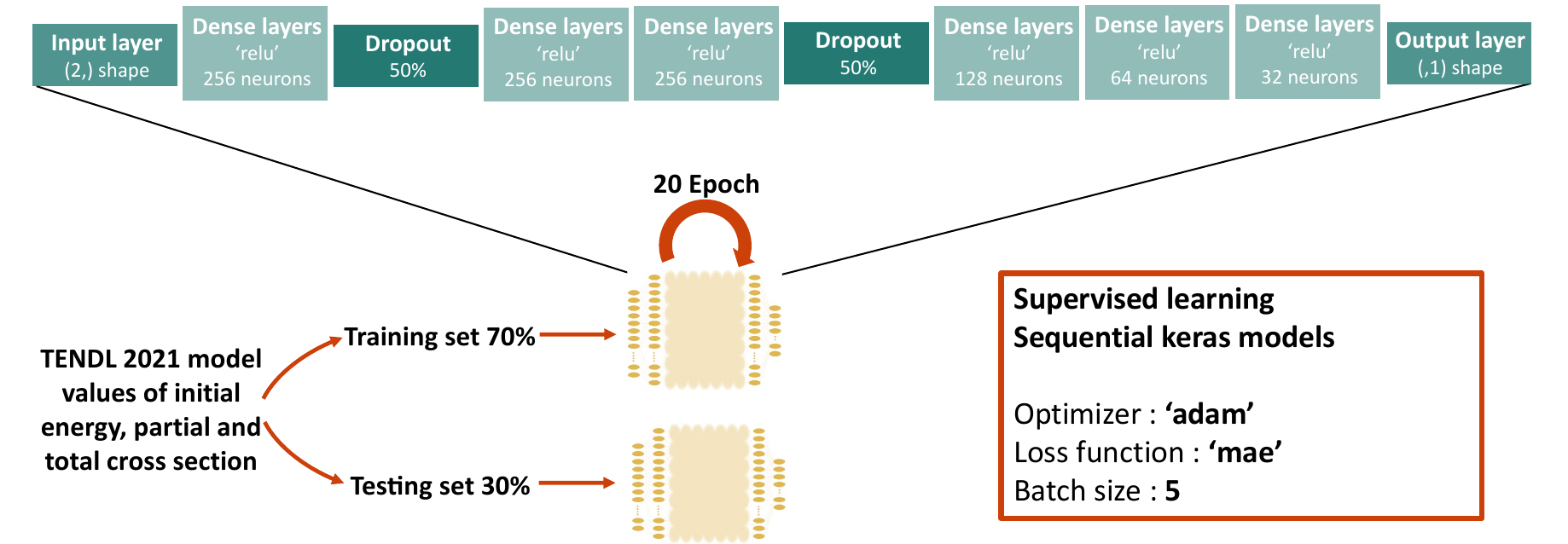}
\caption{\label{fig:DINo_structure} Global structure of the DINo algorithm.}
\end{center}
\end{figure*}

The primary goal of DINo is to predict total cross-sections for initial energies that are not covered by current experimental data. This is crucial in hadrontherapy where accurate treatment plan predictions rely on understanding nuclear reactions and their implications. Figure \ref{fig:DINo_structure} depicts the global structure of the DINo algorithm. The DINo model architecture choices are explained and justified in Annexe of the PhD thesis \cite{Gesson}.

The neural network used in the DINo algorithm is a dense, fully connected, sequential model. The total number of the model parameters is 175 617, all trainable, with a memory usage of 686 kB. This architecture employs dropout layers (with 50$\%$ dropout) to mitigate overfitting by introducing noise and preventing the model from becoming overly dependent
on specific neurons during training. This adds additional variability to the model training process.

During the learning phase of the algorithm, 70$\%$ of the TENDL model values for charge-changing cross-sections, along with their corresponding initial energy values, are used as input data. Simultaneously, the total cross-section values linked to these initial energies are designated as output. This setup enables the algorithm to learn the underlying patterns and relationships between input and output data, forming the basis for its predictive capabilities.

To assess the proficiency of the algorithm, the remaining 30$\%$ of the TENDL model values are reserved for testing purposes at each epoch. This rigorous testing process ensures that the algorithm generalizes well and can make accurate predictions on unseen data. After undergoing 20 epochs of training, the DINo algorithm is ready for application to experimental data. This transition from theoretical modeling to experimental prediction is a critical step in validating the algorithm's effectiveness and real-world applicability.\\

During the second testing phase, on experimental data, the algorithm receives actual experimental values for charge-changing cross-sections along with their associated initial energies. Leveraging the knowledge gained during the learning phase, DINo can then predict the corresponding experimental values for total cross-sections at these specific initial energies. This predictive capability is essential in scenarios where direct experimental measurements are limited or impractical, allowing extrapolation of crucial information for various nuclear processes in simulation for example.

To assess the predictive performance of the DINo deep learning algorithm, its results are systematically compared with the well-established TENDL-2021 model. A simplified $\chi^2$ metric is employed to evaluate the goodness-of-fit between the predicted and experimental cross-sections. The $\chi^2$ value is defined as:
\[
\chi^2 = \sum \frac{(\text{predicted} - \text{expected})^2}{\text{expected}}
\]

For each DINo prediction, the algorithm is trained and executed 100 times to ensure statistical robustness, allowing for the computation of a mean $\chi^2$ value and its associated standard deviation. A lower $\chi^2$ value indicates better agreement with experimental data, demonstrating the model’s ability to replicate nuclear reaction cross-sections accurately.

\section{Results}
\subsection{Detailed Study of $^{11}$C Product Cross-Section}
The first step in evaluating DINo’s performance involves the study of the $^{11}$C isotope, which is of particular interest due to the large dataset available in the EXFOR database. This dataset allows for a rigorous evaluation of DINo’s predictive capabilities.

\begin{figure*}[!htbp]
\begin{center}
\includegraphics[scale=0.55]{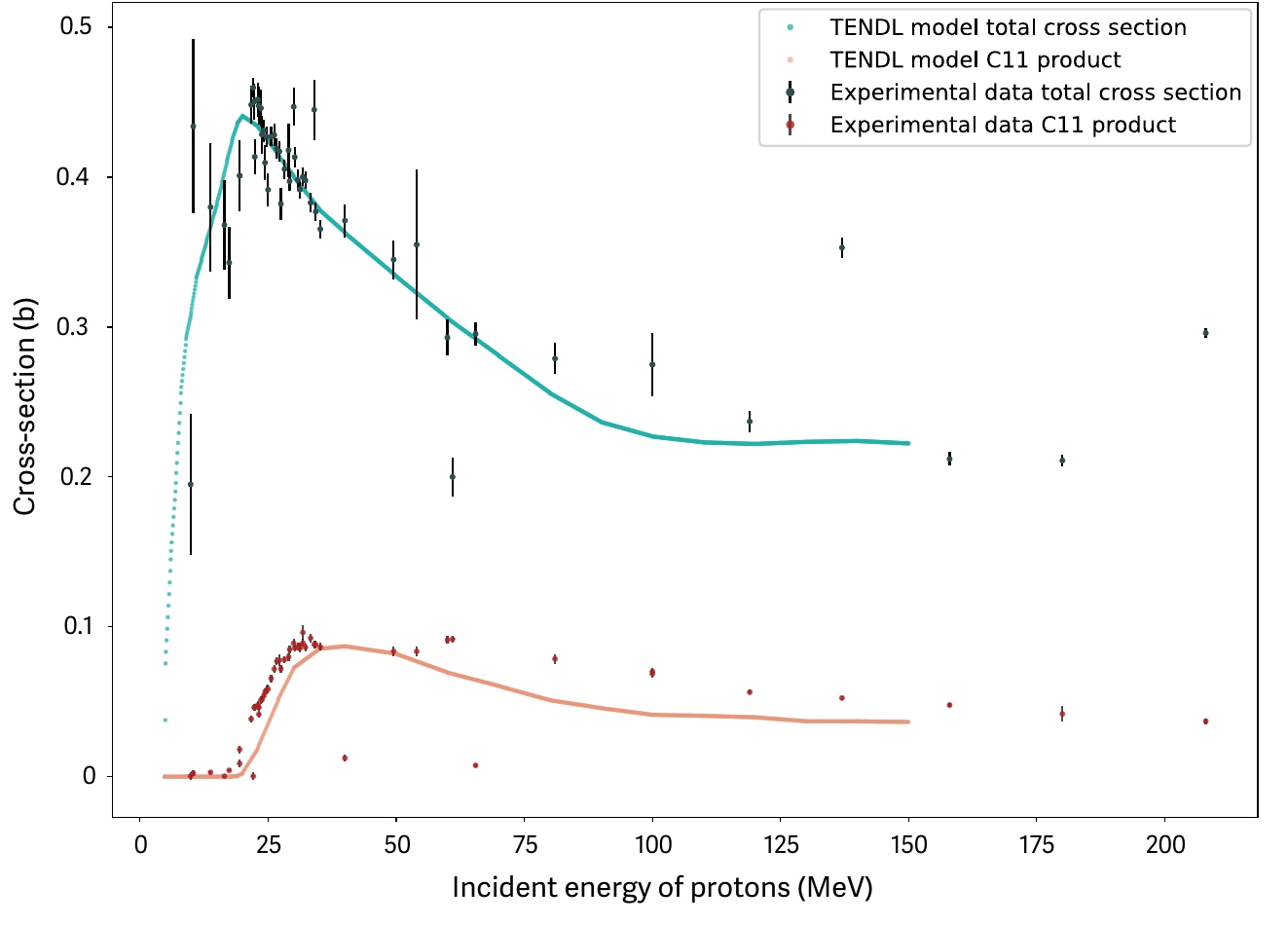}
\caption{\label{fig:TENDLvsData_C11} Total cross-section for incident protons on a $^{12}$C target as a function of incident energy, alongside the $^{11}$C product cross-section: comparison between TENDL-2021 model predictions and experimental data.}
\end{center}
\end{figure*}

Figure \ref{fig:TENDLvsData_C11} presents the total cross-section and $^{11}$C product cross-section for incident protons on a $^{12}$C target, as a function of incident energy. The TENDL-2021 model predictions are plotted alongside experimental data to assess their agreement. The $\chi^2$ evaluation of the TENDL-2021 model’s total cross-section predictions relative to experimental data yields a value of:
\[ \chi^2_{\text{TENDL-2021}} = 7.2 \pm 1.1 \] 
which indicates a quite notable deviation from the experimental dataset.

\begin{figure*}[!htbp]
\begin{center}
\includegraphics[scale=0.55]{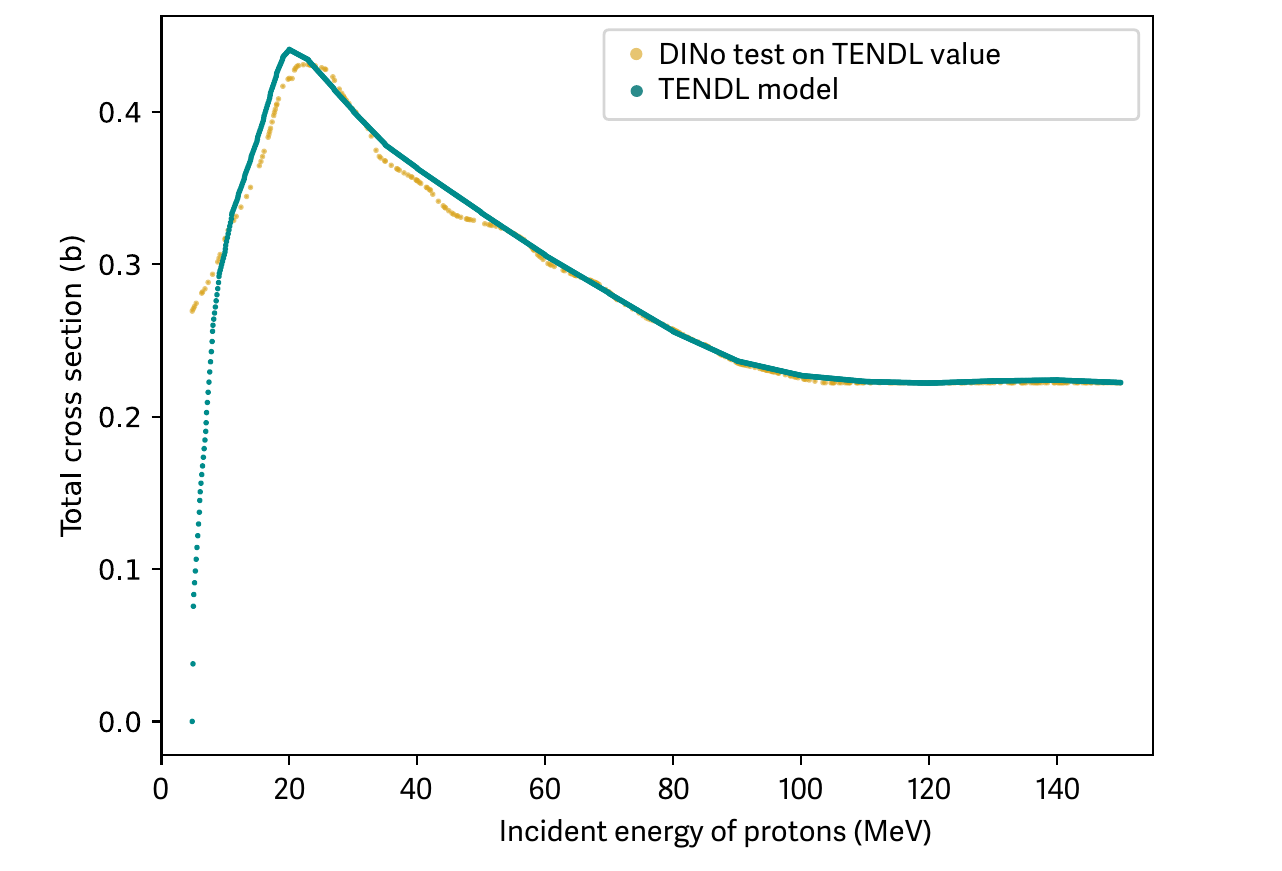}
\caption{\label{fig:total-cross-sections_test} Algorithm test on the total cross-section for incident protons on a $^{12}$C target as a function of incident energy: comparison of TENDL model results with DINo predictions.}
\end{center}
\end{figure*}

To further assess DINo’s ability to replicate nuclear cross-sections, the deep learning algorithm is applied to the same dataset. Figure \ref{fig:total-cross-sections_test} illustrates an iteration of the algorithm tested against the model values. The challenge in deep learning applications lies in ensuring the model generalizes well to unseen data, rather than merely memorizing training examples. It was also verified that the Dino predictions provide better results than a simple scaling of one channel cross-section to another.

\begin{figure*}[!htbp]
\begin{center}
\includegraphics[scale=0.6]{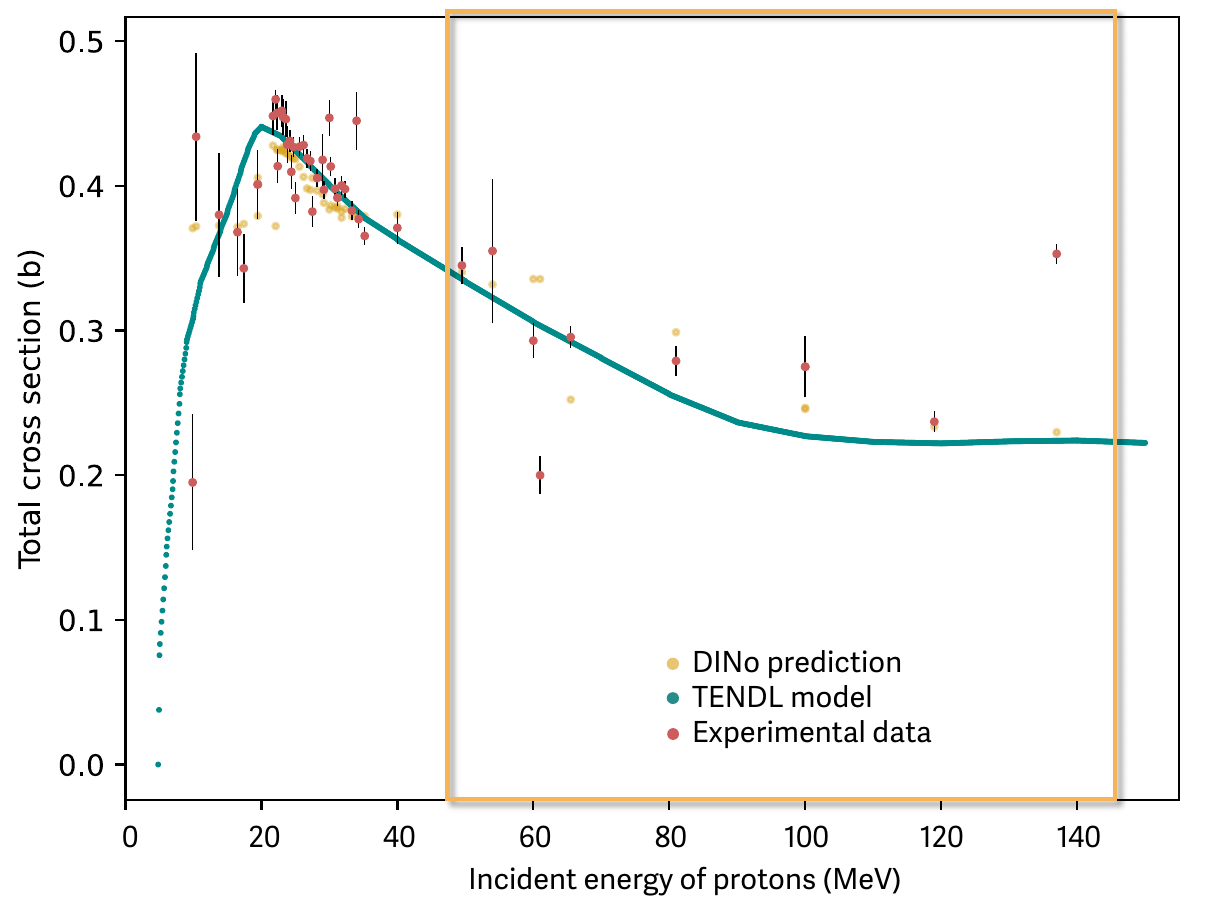}
\caption{\label{fig:total-cross-sections} Total cross-section for incident protons on a $^{12}$C target as a function of incident energy: comparison of TENDL model results, DINo predictions, and experimental data.}
\end{center}
\end{figure*}

A broader comparative analysis between DINo, TENDL-2021, and experimental data is provided in Figure \ref{fig:total-cross-sections}. This visualization highlights the difference in accuracy between both models.
The $\chi^2$ value obtained for DINo’s predictions compared to experimental data is:

\[
\chi^2_{\text{DINo}} = 5.2 \pm 1.0
\]

indicating a significant improvement over the TENDL-2021 model of $\sim 28 \%$ ($\chi^2 = 7.2 \pm 1.1$). Even in the least favorable conditions, when DINo exhibits its lowest predictive performance (maximum $\chi^2 = 6.1$), it still matches or exceeds the best performance of the TENDL-2021 model.

\begin{figure}[htbp]
    \centering
    \includegraphics[width=0.7\textwidth]{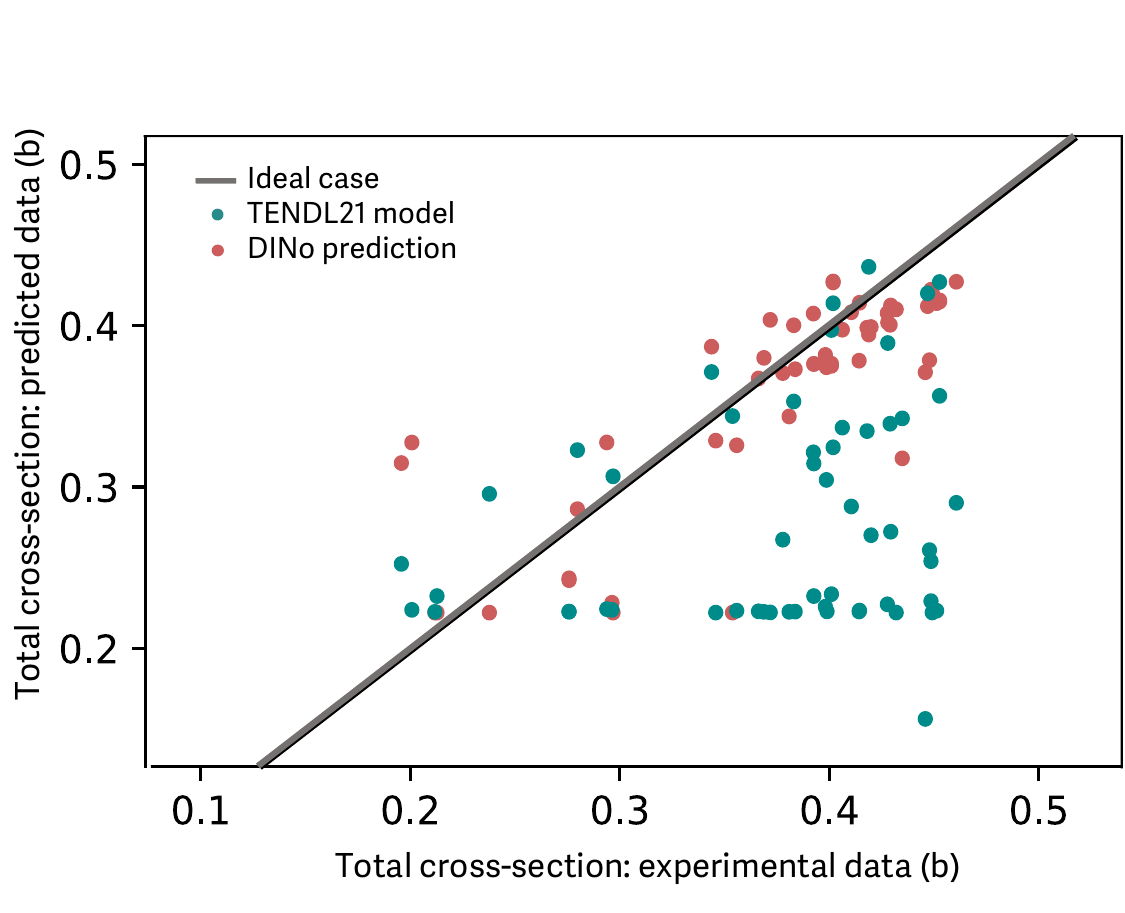}
    \caption{$^{11}$C product cross-sections and total cross-section for incident protons on a $^{12}$C target: comparison between TENDL model results, DINo predictions, and experimental data.}
    \label{fig:dino-tendl-comparison}
\end{figure}

To better visualize the predictive accuracy of both models, Figure \ref{fig:dino-tendl-comparison} presents the predicted total cross-sections for incident protons on a $^{12}$C target, plotted against experimental values. The black linear function represents the ideal case where the prediction equals the experimental value. The closer the data points are to this line, the more accurate the model’s predictions.

\subsection{Extension to other isotopic products}

Beyond $^{11}$C, additional tests were conducted on other charge-changing isotopic products relevant to carbon ion therapy. These isotopes are crucial for accurately modeling proton-carbon interactions, as they influence both the biological and therapeutic effects of hadron therapy.

The selected isotopes are:
\begin{itemize}
    \item \textbf{$^4$He}: alpha particles, are significant due to their high LET, which can contribute to the therapeutic effect but also to potential side effects. 
    \item \textbf{$^6$Li}: Lithium isotopes are important to consider because they can result from the breakup of carbon nuclei and have specific biological impacts.
    \item \textbf{$^9$Be}: Beryllium isotopes are less common but still relevant in understanding fragmentation patterns and dose calculations.
    \item \textbf{$^{10}$B }: Boron isotopes are produced in fragmentation and are of interest, especially considering boron neutron capture therapy (BNCT) applications.
\end{itemize}

Table \ref{tab:chi2} summarizes the $\chi^2$ values comparing DINo’s predictions to TENDL-2021 across these isotopes. Across all isotopes studied, DINo systematically achieves lower $\chi^2$ values, confirming that it predicts outcomes more accurately than TENDL-2021.

\begin{table*}[!hbt]
\begin{center}
	\begin{tabular}{|l||c|c|}
	\hline
    Isotope & DINo's prediction \(\chi^2\) & TENDL-2021 Model \(\chi^2\) \\ \hline
	\hline
    Helium-4 ($^4$He) & 2.2 ± 0.4 & 3.0 ± 0.5 \\
	\hline
    Lithium-6 ($^6$Li) & 0.8 ± 0.3 & 1.0 ± 0.5 \\
    \hline
    Boron-10 ($^{10}$B) & 2.1 ± 0.3 & 2.5 ± 0.7\\
    \hline
    Beryllium-9 ($^9$Be) & 0.7 ± 0.2 & 0.9 ± 0.5 \\ 
    \hline
	\end{tabular}
	\caption{\label{tab:chi2} $\chi^2$ values for charge-changing cross-sections: comparison of DINo predictions with TENDL-2021 model.}
\end{center}
\end{table*}

\section{Discussion}

The results presented in the previous section demonstrate that DINo outperforms TENDL-2021 in predicting nuclear cross-sections, particularly in the high-energy range (above 50 MeV), which is crucial for applications such as particle therapy. The lower $\chi^2$ values obtained for DINo across multiple isotopes confirm the superior predictive capability of deep learning-based approaches over traditional nuclear reaction models.  

The overall trend observed in Figures \ref{fig:TENDLvsData_C11} and \ref{fig:total-cross-sections} suggests that DINo produces predictions that are systematically closer to experimental data than TENDL-2021. The statistical validation in Figure \ref{fig:dino-tendl-comparison} further supports this conclusion, as the majority of DINo predictions cluster near the ideal reference line, indicating a better agreement with experimental values compared to TENDL-2021.\\

The $\chi^2$ values presented in Table \ref{tab:chi2} highlight a consistent improvement across all tested isotopes. The mean $\chi^2$ value of DINo (5.2 ± 1.0) is significantly lower than that of TENDL-2021 (7.2 ± 1.1) for the $^{11}$C cross-section, confirming that DINo provides a better fit to experimental data. Even in the least favorable scenarios, DINo’s worst-case performance (maximum $\chi^2 = 6.1$) matches or exceeds the best performance of TENDL-2021.

Across all charge-changing isotopes studied, DINo achieves a lower $\chi^2$ value than TENDL-2021. This improvement is most pronounced for high-energy cross-sections, which are particularly relevant for medical physics applications such as carbon ion therapy. The superior performance in this range indicates that deep learning methods capture nuclear reaction patterns more accurately than purely theoretical models.\\

A key advantage of DINo over traditional nuclear models like TENDL-2021 is its computational efficiency. Once trained, the deep learning model produces cross-section predictions in the order of microseconds, whereas TENDL relies on extensive simulations, requiring significantly more computational time. This real-time prediction capability of DINo makes it particularly advantageous for applications requiring fast nuclear data processing like hadron therapy treatment planning, or accelerator-based experiments.

Despite its advantages, DINo’s performance remains highly dependent on the quality of training data. The accuracy of predictions deteriorates in energy regions with limited experimental data (e.g., below 10 MeV and above 100 MeV), where the models perform less well and where less data are available for statistical evaluation of the predictions. This observation suggests that:
\begin{itemize}
    \item DINo requires a comprehensive training dataset to maintain accuracy across all energy ranges.
    \item Sparse experimental data limits the model’s generalization capabilities, leading to higher prediction uncertainty.
\end{itemize}

Another limitation observed is the risk of overfitting, particularly in cases where the model is trained on a narrow dataset. In such cases, DINo may memorize noise in the data, leading to reduced accuracy when tested on unseen experimental conditions. To mitigate this issue, future iterations of DINo should integrate data augmentation techniques and regularization strategies to improve generalization performance.

One of the most promising aspects of DINo is its ability to generalize predictions across multiple isotopes, as demonstrated in Table \ref{tab:chi2}. The model maintains high predictive accuracy not just for $^{11}$C, but also for $^4$He, $^6$Li, $^9$Be, and $^{10}$B, with consistently lower $\chi^2$ values than TENDL-2021.

The ability to extrapolate nuclear cross-sections across isotopes suggests that deep learning models like DINo could eventually be used as broad-spectrum nuclear data predictors, offering a complementary tool to traditionally evaluated nuclear data libraries.

\section{Conclusion}
This work has introduced DINo, a deep learning-based approach for improving the prediction of nuclear reaction cross-sections. By leveraging artificial intelligence, DINo demonstrates enhanced predictive accuracy compared to the TENDL-2021 model, particularly in the case of proton interactions with a $^{12}$C target. The model successfully learns correlations between charge-changing and total cross-sections, enabling more accurate predictions even in energy ranges where experimental data are scarce.

The results highlight that DINo systematically outperforms TENDL-2021 across multiple isotopes, as evidenced by lower $\chi^2$ values when compared to experimental data. This improvement is particularly significant for applications such as hadron therapy, where precise cross-section estimations are essential for treatment planning and radiation dose optimization. Moreover, the ability of DINo to generalize across different isotopes suggests its potential for broader applications in nuclear data modeling.

Despite these promising results, certain limitations must be addressed. The performance of DINo remains dependent on the quality and quantity of available experimental data. In regions with sparse data, prediction uncertainties increase, underscoring the need for continuous improvement in training datasets. Additionally, while overfitting was mitigated through regularization techniques, future iterations should further explore data augmentation and other model generalization strategies to enhance robustness.

Going forward, DINo could be extended to predict a wider range of nuclear interactions beyond proton-induced reactions, integrating more complex nuclear reaction channels. Additionally, its integration into Monte Carlo transport codes could further enhance nuclear simulations by providing refined cross-section estimates in real-time.

Finally, this study demonstrates that deep learning can be a powerful complementary tool to traditional nuclear models, offering a data-driven approach to improving nuclear reaction predictions. With continued refinement and validation, such approaches could contribute significantly to nuclear science applications, including reactor physics, and medical physics.

\newpage
\textbf{Data Availability Statement: No Data associated in the manuscript}


\begin{thebibliography}{99}

\bibitem{Battistoni2016} G. Battistoni, I. Mattei, and S. Muraro, "Nuclear physics and particle therapy," \textit{Advances in Physics: X}, vol. 1, no. 4, pp. 661--686, 2016. DOI: \href{https://doi.org/10.1080/23746149.2016.1237310}{10.1080/23746149.2016.1237310}.

\bibitem{TENDL} A. Koning, D. Rochman, J.-C. Sublet, N. Dzysiuk, M. Fleming, and S. van der Marck, "TENDL: Complete nuclear data library for innovative nuclear science and technology," \textit{Nuclear Data Sheets}, vol. 155, pp. 1--55, 2019. DOI: \href{https://doi.org/10.1016/j.nds.2019.01.002}{10.1016/j.nds.2019.01.002}.

\bibitem{ENDF} D.A. Brown, M.B. Chadwick, R. Capote, A.C. Kahler, A. Trkov, M.W. Herman, Y. Danon, A.S. Carlson, et al., "ENDF/B-VIII.0: The 8th Major Release of the Nuclear Reaction Data Library with CIELO-project Cross Sections, New Standards and Thermal Scattering Data," \textit{Nuclear Data Sheets}, vol. 148, pp. 1-142, 2018. DOI: \href{https://doi.org/10.1016/j.nds.2018.02.001}{10.1016/j.nds.2018.02.001}.

\bibitem{FOOT} S. Biondi et al., "The Fragmentation of Target Experiment (FOOT) and its DAQ System," in \textit{IEEE Real Time Conference}, 2020. DOI: \href{https://doi.org/10.48550/arXiv.2010.16251}{10.48550/arXiv.2010.16251}.

v

\bibitem{Gesson} L. Gesson, "Nuclear data for particle therapy," Ph.D. dissertation, Université de Strasbourg, 2024. Available: \url{https://theses.hal.science/tel-04906694}.

\bibitem{ENDF6_manual}
A. Trkov and D. A. Brown, 
``ENDF-6 Formats Manual: Data Formats and Procedures for the Evaluated Nuclear Data Files,'' 
Brookhaven National Laboratory, 2018. DOI: \href{https://doi.org/10.2172/1425114}{10.2172/1425114}.
Available: \url{https://www.osti.gov/biblio/1425114}

\bibitem{EXFOR} International Atomic Energy Agency (IAEA), "Experimental Nuclear Reaction Data (EXFOR) Database," Accessed: 2024. Available: \url{https://www-nds.iaea.org/exfor/}.

\bibitem{Tensorflow} M. Abadi, A. Agarwal, P. Barham, E. Brevdo, Z. Chen, C. Citro et al., "TensorFlow: Large-Scale Machine Learning on Heterogeneous Systems," 2015. Software available from \url{https://www.tensorflow.org/}.

\bibitem{Keras} TensorFlow Developers, "Keras - TensorFlow API," 2024. Available: \url{https://www.tensorflow.org/guide/keras}. Accessed: 2024-02-19.


\end{thebibliography}
\end{document}